\documentclass[onecolumn,showpacs,preprintnumbers]{revtex4}
\usepackage{mathrsfs}
\usepackage{graphicx}
\usepackage{dcolumn}
\usepackage{bm}
\usepackage{epsfig}
\usepackage{amsmath,amssymb}

\begin{document}
\title{The analysis of the charmonium-like states $X^{*}(3860)$,$X(3872)$, $X(3915)$, $X(3930)$ and
$X(3940)$ according to its strong decay behaviors.}
\author{Guo-Liang Yu$^{1}$}
\email{yuguoliang2011@163.com}
\author{Zhi-Gang Wang$^{1}$}
\email{zgwang@aliyun.com}
\author{Zhen-Yu Li$^{2}$}

\affiliation{$^1$ Department of Mathematics and Physics, North China
Electric power university, Baoding 071003, People's Republic of
China \\$^2$ School of Physics and Electronic Science, Guizhou
Normal College, Guiyang 550018, People's Republic of China}
\date{\today }

\begin{abstract}
Inspired by the newly observed state $X^{*}(3860)$, we analyze the
strong decay behaviors of some charmonium-like states
$X^{*}(3860)$,$X(3872)$, $X(3915)$, $X(3930)$ and $X(3940)$ by the
$^{3}P_{0}$ model. We carry out our work based on the hypothesis
that these states are all being the charmonium systems. Our analysis
indicates that $0^{++}$ charmonium state can be a good candidate for
$X^{*}(3860)$ and $1^{++}$ state is the possible assignment for
$X(3872)$. Considering as the $3^{1}S_{0}$ state, the decay behavior
of $X(3940)$ is inconsistent with the experimental data. So, we can
not assign $X(3940)$ as the $3^{1}S_{0}$ charmonium state by present
work. Besides, our analysis implies that it is reasonable to assign
$X(3915)$ and $X(3930)$ to be the same state, $2^{++}$. However,
combining our analysis with that of Zhou~\cite{ZhouZY}, we speculate
that $X(3915)$/$X(3930)$ might not be a pure $c\overline{c}$
systems.
\end{abstract}


\maketitle

\begin{large}
\textbf{1 Introduction}
\end{large}

Very recently, the Belle Collaboration observed a new
charmonium-like state, $X^{*}(3860)$, by performing a full amplitude
analysis of the process $e^{+}e^{-}\rightarrow J/\psi
D\overline{D}$~\cite{Bell1}. Its mass is
$(3862^{+26+40}_{-32-13}MeV/c^{2})$ and width is
$(201^{+154+88}_{-67-82})MeV$. The $J^{PC}=0^{++}$ hypothesis is
favored over the $2^{++}$ assignment at the level of $2.5\sigma$.
In reference~\cite{WangZG6}, this state was explained to be
a $C\gamma_{5}\bigotimes \gamma_{5} C$ type scalar tetraquark state
by the method of QCD sum rules(QCDSR). Actually, people once had assigned
$X(3915)$ as the $0^{++}$ charmnium state~\cite{LiuX1,Lees1}, which
was observed by Belle, BABAR Collaboration in $B\rightarrow J/\psi\omega K$ decay
mode~\cite{Choi1,Amo,Uehara,Lees1,Abe}. Its mass and width are
listed in Table I.

After $X(3915)$ was suggested to be the $\chi_{c0}$ assignment, it
encountered several challenges~\cite{Guo,Olsen2,Eichten,WangZG4}.
For example, the decay $\chi_{c0}(2P)\rightarrow D\overline{D}$,
which was expected to be the dominant decay mode, has not been
observed experimentally. In contrast, the decay mode
$X(3915)\rightarrow J\psi \omega$, which should be
OZI(Okubo-Zweig-Iizuka)~\cite{Okubo} suppressed, was observed
instead in experiments. In addition, the mass splitting of
$\chi_{c2}(2P)-\chi_{c0}(2P)$ is too small. A reanalysis of the data
from Ref.~\cite{Lees1}, presented in Ref.~\cite{ZhouZY}, showed that
$X(3915)$ could be the same state as $X(3930)$, whose quantum number
is $2^{++}$~\cite{Beringer}, due to the degeneracies of their masses
and widths. Now, the observation of $X^{*}(3860)$ ,which was
assigned to be the $0^{++}$ state, is an important verification of
the results in Ref.~\cite{ZhouZY}.

The mass of this newly observed $X^{*}(3860)$ is close to that of
another charmonium-like state $X(3872)$. However, these two hadrons
are impossible to be the same state because of its different decay
modes and widths(see Table I). After $X(3872)$ was discovered by
Belle Collaboration~\cite{Choi2} and confirmed by
BABAR~\cite{Aubert}, CDF~\cite{Acosta}, D0~\cite{Abazov} and
Bell~\cite{Gokhroo} Collaborations, its nature has still been very
controversial. It was mainly explained to be such structures as a
molecule
state~\cite{Close,Voloshin,Wong,Swanson,Tornqvist,Mohammad,Larionov,WangZG1,Esposito,KangXW},
a hybrid charmonium~\cite{LiBA,Nielsen,Takizawa}, a tetraquark
state~\cite{CuiY,Matheus,Chiu,Dubnicka,WangZG2}. Another important
explanation is that it was the charmonium state with quantum of
$1^{++}$~\cite{Achasov,Deng}, which has a dominant decay mode
$D^{0}\overline{D}^{*0}$.

Belle Collaboration reported another charmonium-like state $X(3940)$
from the inclusive process  $e^{+}x^{-}\rightarrow J/\psi+$cc at a
mass of $M=(3.943\pm0.006\pm0.006)$GeV/$c^{2}$~\cite{Abe2}. Later,
its decay width was confirmed to be
$\Gamma=(37^{+26}_{-15}\pm8)$MeV~\cite{Pakhlov}. People have also
explored the structure of $X(3940)$ with different kinds of methods
such as the light-cone formalism~\cite{Braguta}, the NRQCD
factorization formula~\cite{ZhuRL,HeZG} and
QCDSR~\cite{Albuquerque,WangZG3}. According to these studies, there
seems to be no doubt that the quantum number of $X(3940)$ is
$3^{1}S_{0}$. However, its structure is still controversial, which
have been explained to be different states such as the charmonium
state~\cite{Braguta}, the molecular
state~\cite{Albuquerque,WangZG3,LiuX2} and a Mixed
Charmonium-Molecule State~\cite{Albuquerque2,Fernandez}.
\begin{table*}[htbp]
\begin{ruledtabular}\caption{The experimental information about the $X$ states in this paper.}
\begin{tabular}{c c c c c c }
States & \  Mass(MeV/c$^{2}$)  & \ Width(MeV)  & \ $J^{PC}$ &\ Decay channels  \\
\hline
X$^{*}$(3860)~\cite{Bell1} & \  $3862^{+26+40}_{-32-13}$         &  \ $201^{+154+88}_{-67-82}$     & \  $0^{++}(2^{3}P_{0})$   &  \   $D\overline{D}$   \\
\hline
X(3915)~\cite{Lees1} & \   $3919.4\pm2.2\pm1.6$         &  \ $13\pm6\pm3$     & \  $0^{++}(2^{3}P_{0})$,$2^{++}(2^{3}P_{2})$    &  \   $J/\psi\omega$  \\
\hline
X(3930) & \   $3929\pm5\pm2$~\cite{Uehara2}        &  \ $29\pm10\pm2$    & \  $2^{++}(2^{3}P_{2})$    &  \   $D\overline{D}$  \\
        &\    $3926.7\pm2.7\pm1.1$~\cite{Aubert4}  &\   $21.3\pm6.8\pm3.6$ &\  $2^{++}(2^{3}P_{2})$  &\ $D\overline{D}$ \\
\hline
     & \   $3872\pm0.6\pm0.5$~\cite{Choi2}        &  \        & \      &  \   $J/\psi\pi^{+}\pi^{-}$   \\
       & \   $3871.3\pm0.7\pm0.4$~\cite{Acosta}        &  \     & \      &  \   $J/\psi\pi^{+}\pi^{-}$  \\
X(3872)  & \   $3871.8\pm3.1\pm3.0$ ~\cite{Abazov}       &  \  $<2.3$   & \   $1^{++}(2^{3}P_{1})$   &  \   $J/\psi\pi^{+}\pi^{-}$ \\
     & \   $3873.4\pm1.4$~\cite{Aubert}        &  \      & \      &  \   $J/\psi\pi^{+}\pi^{-}$   \\
      & \   $3875.4\pm0.7^{+1.2}_{-2.0}$~\cite{Gokhroo}        &  \      & \      &  \   $D^{0}\overline{D}^{0}\pi^{0}$,$J/\psi\omega$    \\
      & \   $3875.6\pm0.7^{+1.4}_{-1.5}$~\cite{Grenier5}        &  \      & \      &  \   $D^{0}\overline{D}^{*0}+h.c.$,$J/\psi\rho$    \\
\hline
X(3940)~\cite{Abe2,Pakhlov} & \   $3942^{+7}_{-6}\pm6$          &  \ $37^{+26}_{-15}\pm8$     & \  $0^{-+}(3^{1}S_{0})$    &  \   $D\overline{D}^{*}$   \\
\end{tabular}
\end{ruledtabular}
\end{table*}

In summary, these newly discovered charmonium-like states have
inspired many interests about their phyisical natures. In order to
further study its structures, we perform an analysis of the strong
decay behaviors of $X^{*}(3860)$, $X(3872)$, $X(3915)$, $X(3930)$
and $X(3940)$ with the $^{3}P_{0}$ decay model. The experimental
information about these states are listed in Table I. Since these
$X$ states can not be completely ruled out from the $c\overline{c}$
systems at present, we carry out our calculations by assuming them
to be the charmoniums. Our analysis will be helpful to confirm or
exclude some $c\overline{c}$ systems and useful to further determine
the quantum numbers of the confirmed charmonium states. As for the
strong decays of the hadrons, $^{3}P_{0}$ decay
model~\cite{Micu,Carlitz,Yaouanc} is an effective method. It has
been widely used in this field since it gives a good description of
the decay behaviors of many
hadrons~\cite{Blunder,ZhouHQ,LiDM,ZhanbB,Ackleh,Close3,Ferretti,GuoLY}.
The article is arranged as follows: In section 2, we give a brief
review of the $^{3}P_{0}$ decay model; in Sec.3 we study the strong
decays of $X^{*}(3860)$, $X(3872)$, $X(3915)$, $X(3930)$ and
$X(3940)$; in Sec.4, we present our conclusions.

\begin{large}
\textbf{2 The decay model}
\end{large}
\begin{figure}[h]
  \includegraphics[width=15cm]{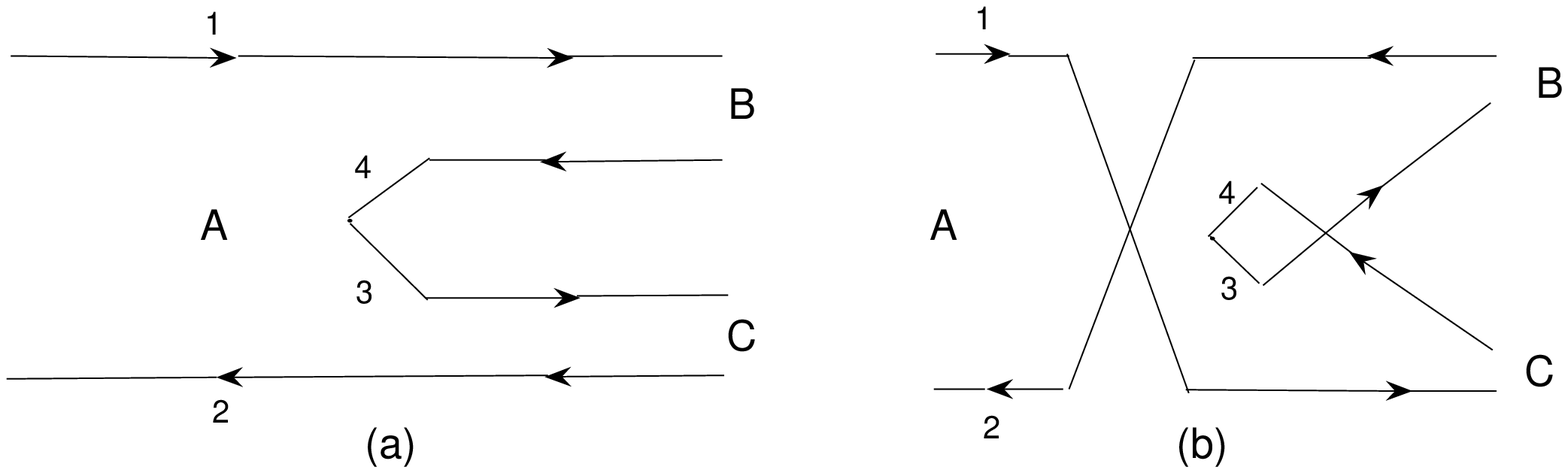}
  \caption{The two possible diagrams contributing to $A\rightarrow BC$ in the $^{3}P_{0}$ model.}\label{Figure 1:}
\end{figure}

The principle of $^{3}P_{0}$ decay model is illustrated clearly in
Fig.1, where a quark-antiquark pair($q_{3}\overline{q}_{4}$) is
created from the vacuum with $0^{++}$ quantum numbers. With the
$q_{1}\overline{q}_{2}$ within the initial meson, this quark systems
regroups into two outgoing mesons via quark rearrangement for the
meson decay process $A$$\rightarrow$$BC$. Its transition operator in
the nonrelativistic limit reads
\begin{equation}
\begin{split}
    T=
    &-3\gamma\sum_{m}\langle1m1-m\mid00\rangle\int d^{3}\vec p_{3}d^{3}\vec p_{4}\delta^{3}(\vec p_{3}+\vec p_{4})\mathcal{Y}_{1}^{m}(\frac{\vec p_{3}-\vec p_{4}}{2})
    \chi_{1-m}^{34}\varphi_{0}^{34}\omega_{0}^{34}b_{3}^{\dag}(\vec p_{3})d_{4}^{\dag}(\vec p_{4})
\end{split}
\end{equation}
where $\gamma$ is a dimensionless parameter reflecting the
creation strength of the quark-antiquark $q_{3}\overline{q}_{4}$
pair. The solid harmonic polynomial
$\mathcal{Y}_{1}^{m}(\vec p)\equiv|\vec
p|^{1}Y_{1}^{m}(\theta_{p},\phi_{p})$ reflects the momentum-space
distribution of the $q_{3}\overline{q}_{4}$.

The helicity amplitude of the decay process $A\rightarrow BC$ in the
parent meson $A$ center of mass frame is
\begin{equation}
\begin{aligned}
\mathcal{M}^{M_{J_{A}}M_{J_{B}}M_{J_{C}}}(\vec P)=
&\gamma\sqrt{8E_{A}E_{B}E_{C}}\sum_{\mbox{\tiny$\begin{array}{c}
M_{L_{A}},M_{S_{A}},\\
M_{L_{B}},M_{S_{B}},\\
M_{L_{C}},M_{S_{C}},m\end{array}$}}\langle
L_{A}M_{L_{A}}S_{A}M_{S_{A}}\mid J_{A}M_{J_{A}}\rangle \langle
L_{B}M_{L_{B}}S_{B}M_{S_{B}}\mid J_{B}M_{J_{B}}\rangle \\
&\times\langle L_{C}M_{L_{C}}S_{C}M_{S_{C}}\mid
J_{C}M_{J_{C}}\rangle\langle 1m1-m\mid 00\rangle\langle \chi_{S_{B}M_{S_{B}}}^{14}\chi_{S_{C}M_{S_{C}}}^{32}\mid \chi_{S_{A}M_{S_{A}}}^{12}\chi_{1-m}^{34}\rangle \\
&\times[\langle \phi_{B}^{14}\phi_{C}^{32}\mid \phi_{A}^{12}\phi_{0}^{34}\rangle I(\vec P,m_{1},m_{2},m_{3}) \\
&+(-1)^{1+S_{A}+S_{B}+S_{C}}\langle \phi_{B}^{32}\phi_{C}^{14}\mid \phi_{A}^{12}\phi_{0}^{34}\rangle I(-\vec P,m_{2},m_{1},m_{3})],
\end{aligned}
\end{equation}

where $I(\vec P,m_{1},m_{2},m_{3})$ is the spatial integral which is defined as
\begin{equation}
\begin{split}
I(\vec P,m_{1},m_{2},m_{3})=
&\int d^{3}\vec p \psi^{*}_{n_{B}L_{B}M_{L_{B}}}(\frac{m_{3}}{m_{1}+m_{2}}\vec P_{B}+\vec p)\psi^{*}_{n_{C}L_{C}M_{L_{C}}}(\frac{m_{3}}{m_{2}+m_{3}}\vec P_{B}+\vec p) \\
&\times\psi_{n_{A}L_{A}M_{L_{A}}}(\vec P_{B}+\vec p)\mathcal{Y}_{1}^{m}(\vec p)
\end{split}
\end{equation}
where $\vec P =\vec P_{B} =-\vec P_{C}, \vec p = \vec p_{3}$,
$m_{3}$ is the mass of the created quark $q_{3}$. We employ the
simple harmonic oscillator (SHO) approximation as the meson space
wave functions in Eq.(3).
\begin{equation}
\begin{split}
\Psi_{nLM_{L}}(\vec p)=
&(-1)^{n}(-i)^{L}R^{L+\frac{3}{2}}\sqrt{\frac{2n!}{\Gamma(n+L+\frac{3}{2})}}exp(-\frac{R^{2}p^{2}}{2})L_{n}^{L+\frac{1}{2}}(R^{2}p^{2})\mathcal{Y}_{LM_{L}}(\vec
p)
\end{split}
\end{equation}
Where $R$ is the scale parameter of the SHO. With the Jacob-Wick
formula, the helicity amplitude can be converted into the partial
wave amplitude
\begin{equation}
\begin{split}
\mathcal{M}^{JL}(\vec P)=
&\frac{\sqrt{4\pi(2L+1)}}{2J_{A}+1}\sum_{M_{J_{B}}M_{J_{C}}}\langle
L0JM_{J_{A}}|J_{A}M_{J_{A}}\rangle \langle
J_{B}M_{J_{B}}J_{C}M_{J_{C}}|JM_{J_{A}}\rangle\mathcal{M}^{M_{J_{A}}M_{J_{B}}M_{J_{C}}}(\vec
P)
\end{split}
\end{equation}
where $M_{J_{A}}=M_{J_{B}}+M_{J_{C}}$, $\mathbf{J_{A}=J_{B}+J_{C}}$
and $\mathbf{J_{A}+J_{P}=J_{B}+J_{C}+J_{L}}$. Finally, the decay width in
terms of partial wave amplitudes is
\begin{equation}
\Gamma=\frac{\pi}{4}\frac{|\vec P|}{M_{A}^{2}}\sum_{JL}|\mathcal{M}^{JL}|^{2}
\end{equation}
where $P=|\vec P|=\frac{\sqrt{[M_{A}^{2}-(M_{B}+M_{C})^{2}][M_{A}^{2}-(M_{B}-M_{C})^{2}]}}{2M_{A}}$, $M_{A}$, $M_{B}$, and $M_{C}$ are the masses of the meson
$A$, $B$, and $C$, respectively.

\begin{large}
\textbf{3 The results and discussions}
\end{large}

The decay width based on $^{3}P_{0}$ model depends on the following
input parameters, the light quark pair($q\overline{q}$) creation
strength $\gamma$, the SHO wave function scale parameter $R$, and
the masses of the mesons and the constituent quarks. The adopted
masses of the hadrons are listed in TABLE II, and $m_{u} = m_{d} =
0.22$ GeV, $m_{s} = 0.419$ GeV and $m_{c} = 1.65$
GeV~\cite{Patrignani}.
\begin{table*}[htbp]
\begin{ruledtabular}\caption{The adopted masses of the hadrons used in our calculations.}
\begin{tabular}{c c c c c c c c c c c c c c c c c c c c c}
States & \  $M_{X^{*}(3860)}$  & \ $M_{X(3872)}$  & \ $M_{X(3915)}$ &\ $M_{X(3930)}$ & \ $M_{X(3940)}$  \\
Mass(MeV) & \   3862~\cite{Bell1}        &  \ 3872~\cite{Choi2}     & \  3919~\cite{Lees1}    &  \   3927~\cite{Aubert4}   & \  3942~\cite{Abe2,Pakhlov}    \\
\hline States &\  $M_{D^{\pm}}$ & \ $M_{D^{0}}$ & \ $M_{D^{*\pm}}$
 &\ $M_{D^{*0}}$  &\ \\
 Mass(MeV)~\cite{Patrignani} &  \  1869.6      &  \  1864.83    &  \    2010 & \ 2007
 \\
\end{tabular}
\end{ruledtabular}
\end{table*}
As for the scale parameter $R$, there are mainly two kinds of
choices which are the common value and the effective value. The
effective value can be fixed to reproduce the realistic root mean
square radius by solving the Schrodinger equation with the linear
potential~\cite{Godfrey,LiBQ5}. For the $c\overline{c}$ systems, the
$R$ value of $2P$ states is estimated to be
$2.3\sim2.5$GeV$^{-1}$~\cite{YangYC}. For the mesons $D$ and$D^{*}$,
its value is taken to be $R_{D^{0}[D^{\pm}]}=1.52$GeV$^{-1}$,
$R_{D^{*0}[D^{*\pm}]}=1.85$GeV$^{-1}$~\cite{YangYC,Godfrey} in this
work. Finally, we choose the value of $\gamma$ to be 6.25 for the
creation of the u/d quark following Ref.~\cite{Blunder}.

We know that $X^{*}(3860)$ was favored to be the
$0^{++}(2^{3}P_{0})$ charmonium-like state by Bell Collaboration and
$X(3915)$ had also once been explained to be this assignment.
Lately, the latter one was corrected to be the same state as another
charmonium-like state, $X(3930)$ which had been determined to be
$2^{++}(2^{3}P_{2})$ assignment. In order to further confirm these
conclusions, we study the strong decay behaviors of $X^{*}(3860)$ by
considering it as the $2^{3}P_{0}$ and $2^{3}P_{2}$ charmoniums. And
so does for the $X(3915)$ state. Besides, we also perform an
analysis of the decay behaviors of $X(3930)$, $X(3872)$ and
$X(3940)$ which have been favored to be $2^{++}(2^{3}P_{2})$,
$1^{++}(2^{3}P_{1})$ and $0^{-+}(3^{1}S_{0})$ states, respectively.
As mentioned in Ref.~\cite{Guo}, the mass difference
$M_{X(3930)}-M_{X(3915)}=9.7\pm3.7$ MeV, is smaller than the fine
splitting of $1P$ states
$M_{\chi_{c2}}-M_{\chi_{c0}}=141.45\pm0.32$MeV~\cite{Beringer}. This
is an important evidence to recognizing $X(3915)$ and $X(3930)$ as
the same state. In order to determine its mass precisely, we also
calculate the decay widths of $2^{3}P_{2}(\chi_{c2})$ state on
different masses. All of the results are illustrated in the form of
graphs, which can be seen from Figures 2 to 9.
\begin{figure}[h]
\begin{minipage}[t]{0.45\linewidth}
\centering
\includegraphics[height=5cm,width=7cm]{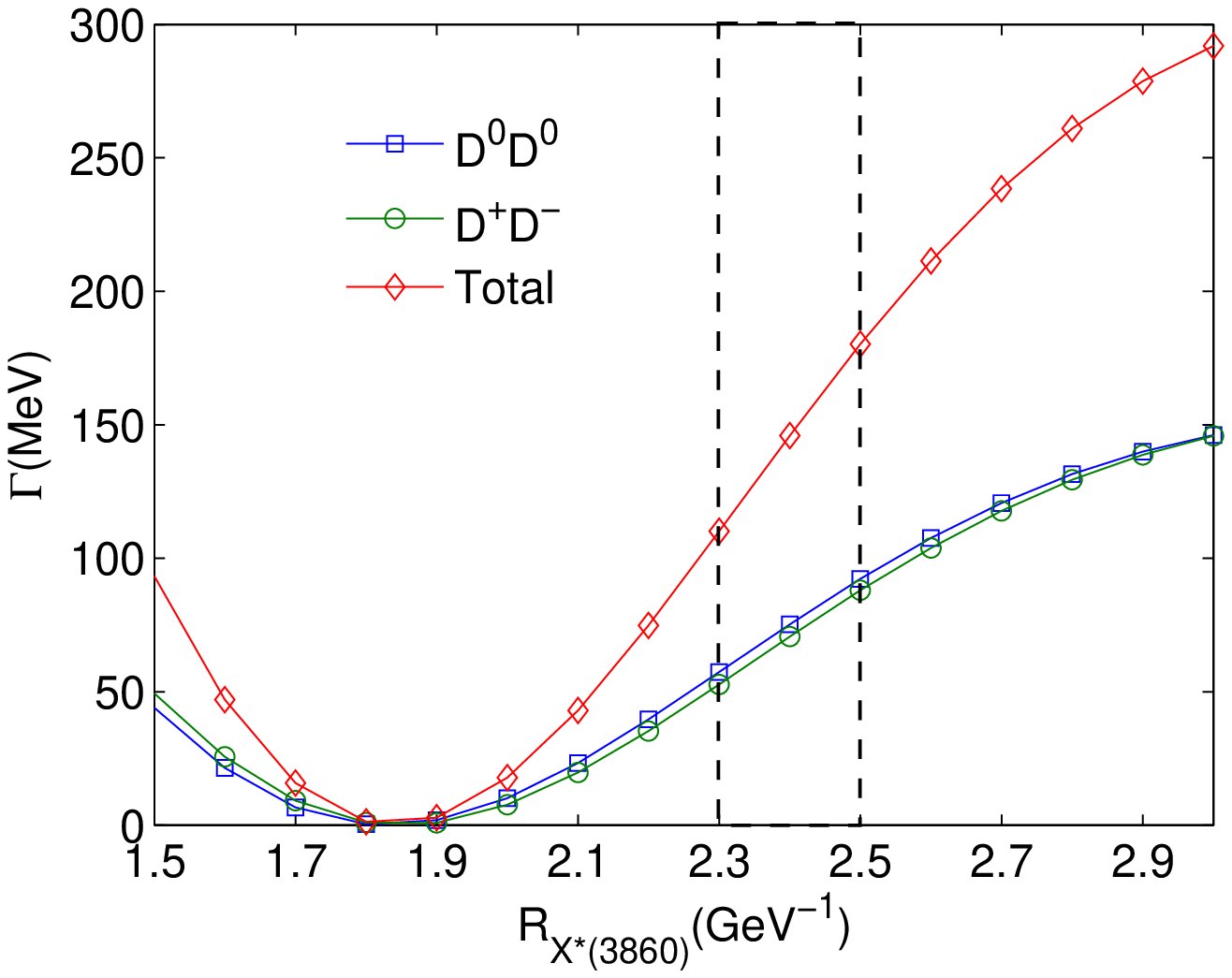}
\caption{The strong decay of $X^{*}(3860)$ as the
$0^{++}(2^{3}P_{0})$ state on scale parameter
$R_{X^{*}(3860)}$.\label{your label}}
\end{minipage}
\hfill
\begin{minipage}[t]{0.45\linewidth}
\centering
\includegraphics[height=5cm,width=7cm]{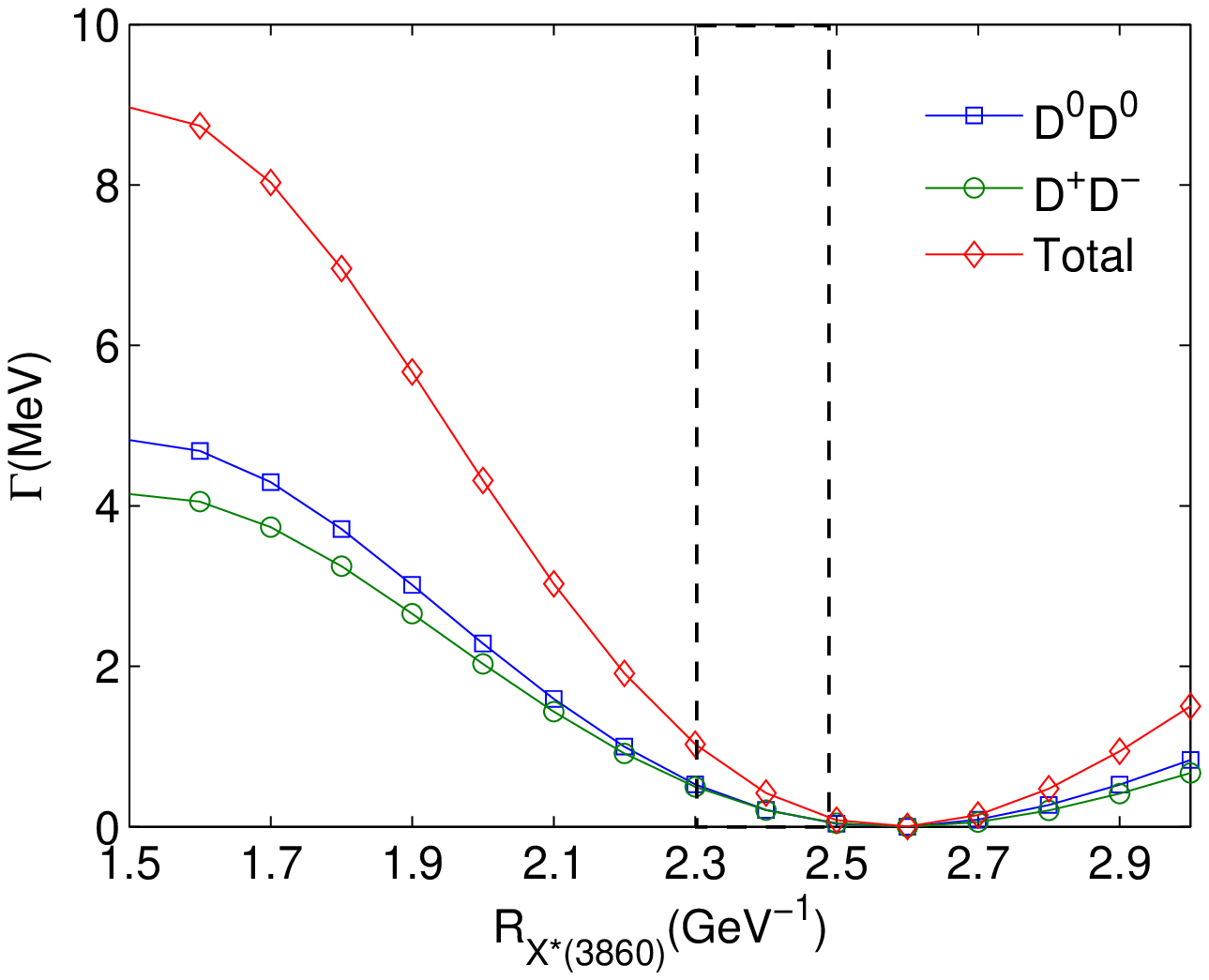}
\caption{The strong decay of $X^{*}(3860)$ as the
$2^{++}(2^{3}P_{2})$ state on scale parameter
$R_{X^{*}(3860)}$.\label{your label}}
\end{minipage}
\end{figure}

\begin{figure}[h]
\begin{minipage}[t]{0.45\linewidth}
\centering
\includegraphics[height=5cm,width=7cm]{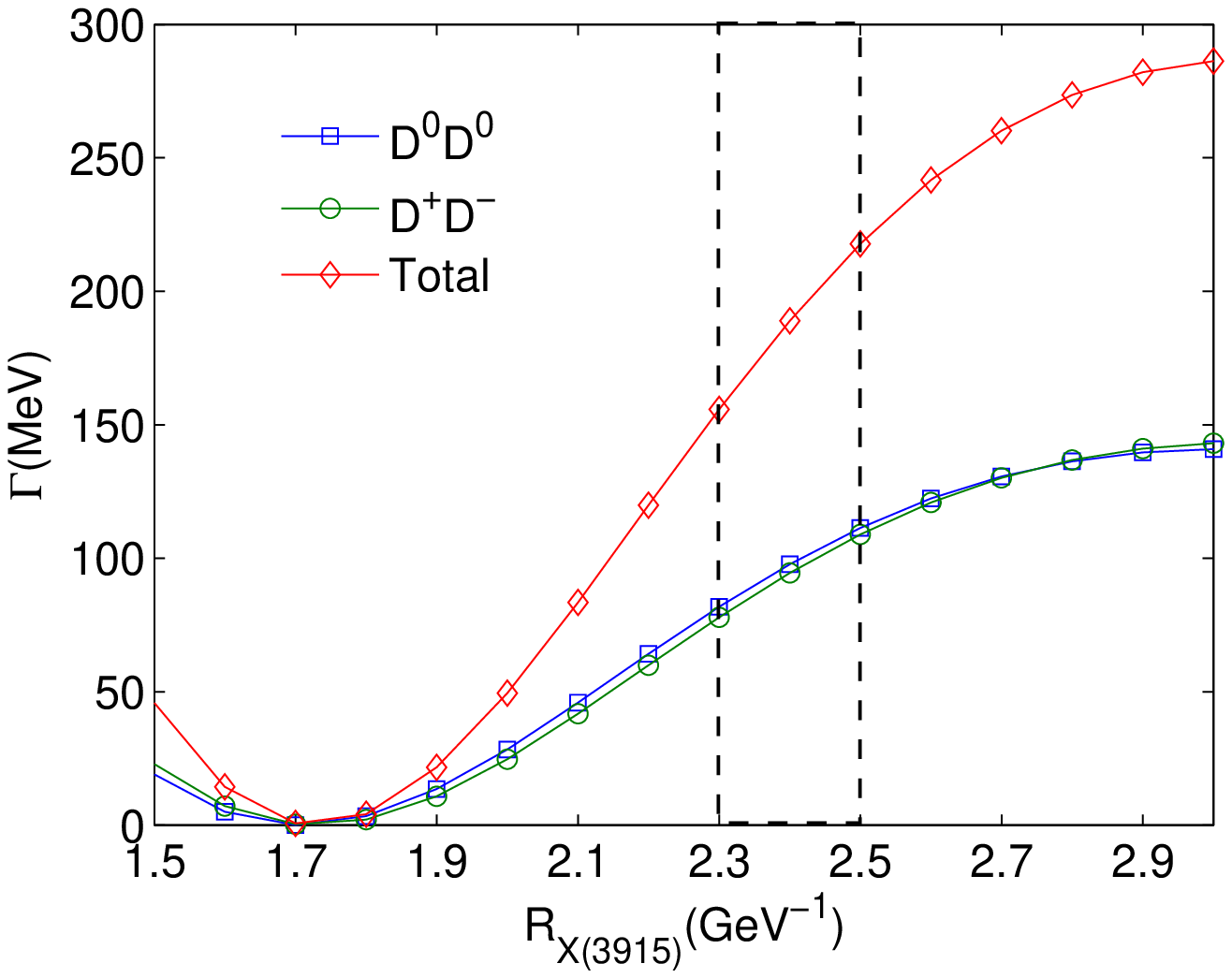}
\caption{The strong decay of $X(3915)$ as the $0^{++}(2^{3}P_{0})$
state on scale parameter $R_{X(3915)}$.\label{your label}}
\end{minipage}
\hfill
\begin{minipage}[t]{0.45\linewidth}
\centering
\includegraphics[height=5cm,width=7cm]{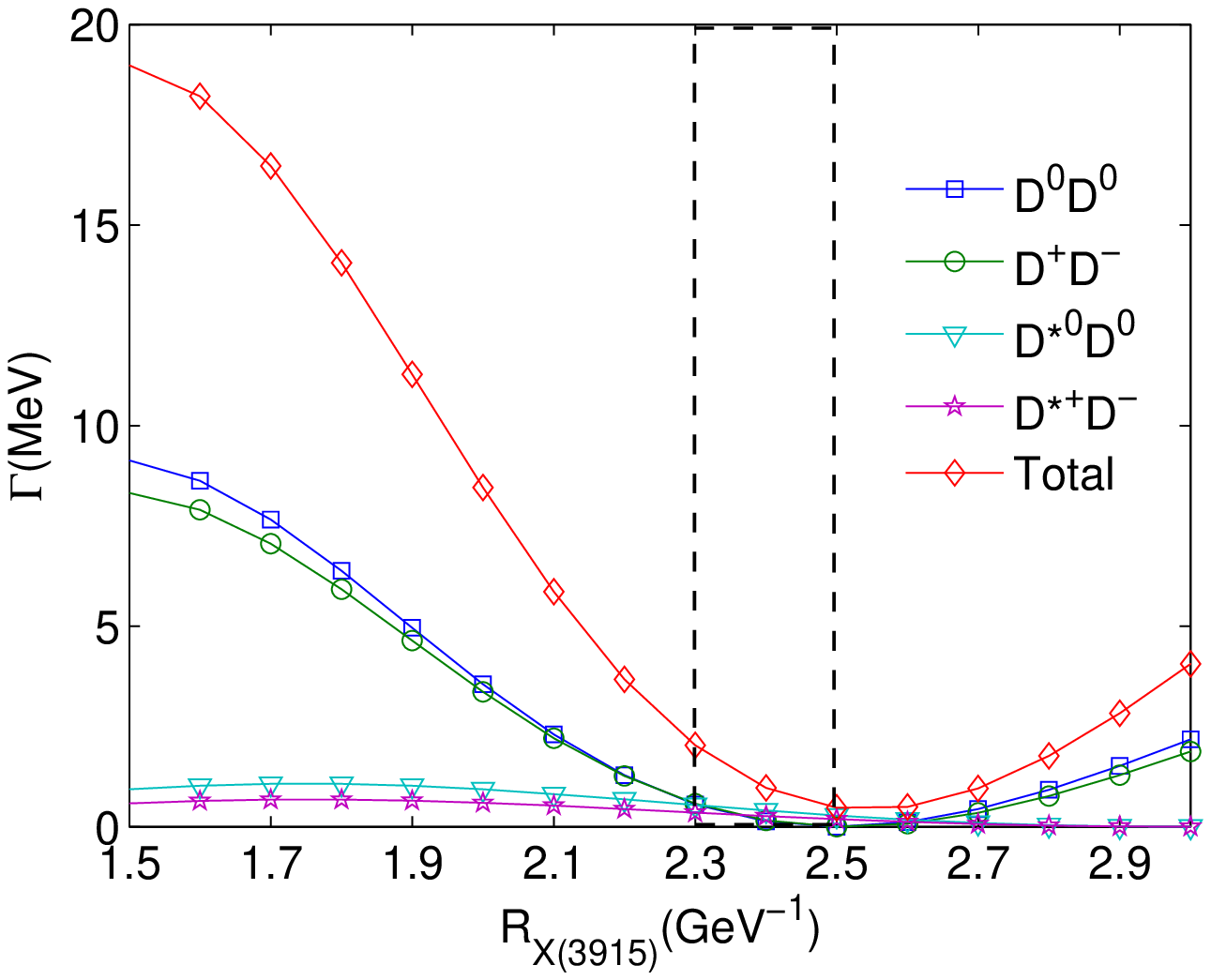}
\caption{The strong decay of $X(3915)$ as the $2^{++}(2^{3}P_{2})$
state on scale parameter $R_{X(3915)}$.\label{your label}}
\end{minipage}
\end{figure}
\begin{figure}[h]
\begin{minipage}[t]{0.45\linewidth}
\centering
\includegraphics[height=5cm,width=7cm]{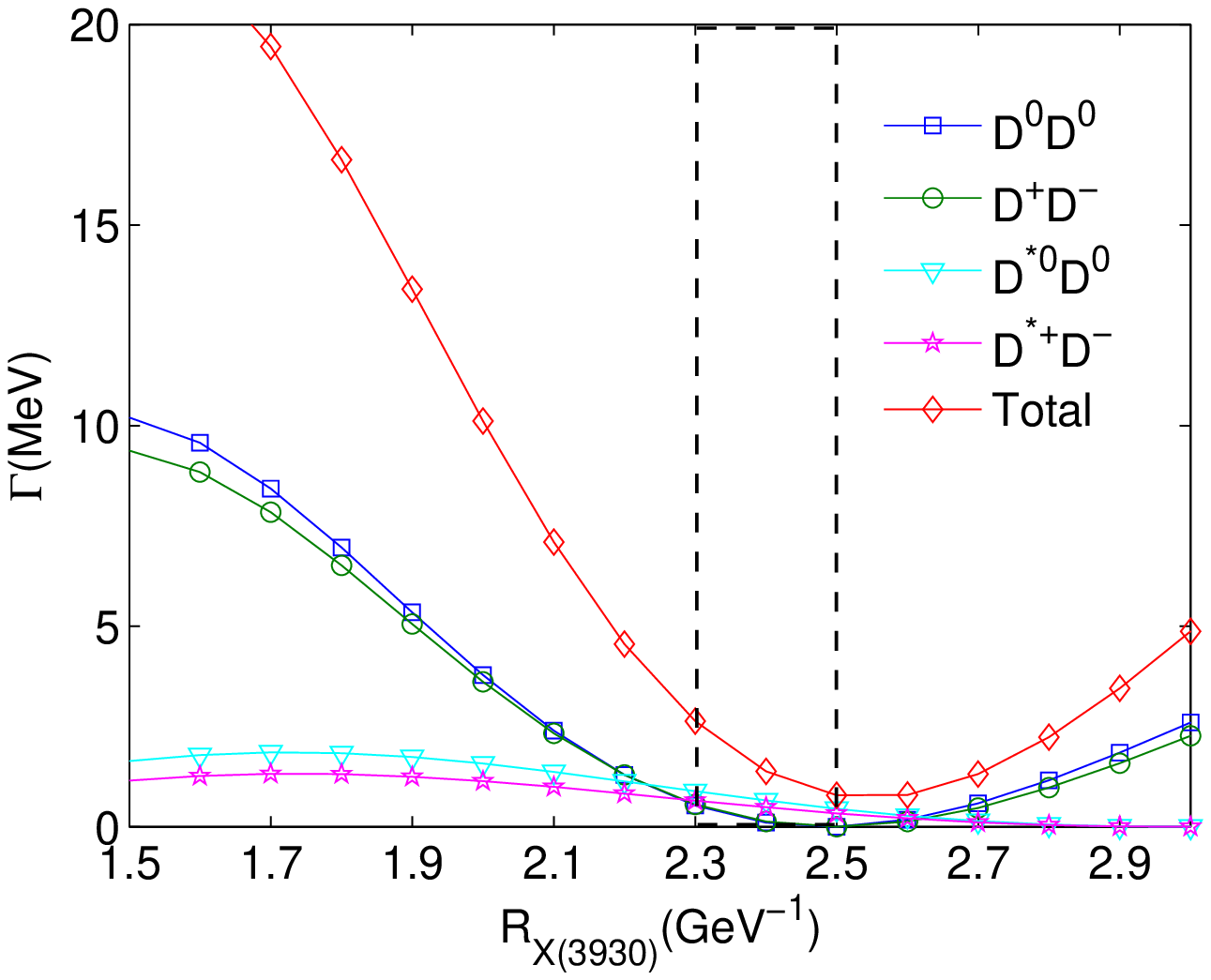}
\caption{The strong decay of $X(3927)$ as the $2^{++}(2^{3}P_{2})$
state on scale parameter $R_{X(3927)}$.\label{your label}}
\end{minipage}
\hfill
\begin{minipage}[t]{0.45\linewidth}
\centering
\includegraphics[height=5cm,width=7cm]{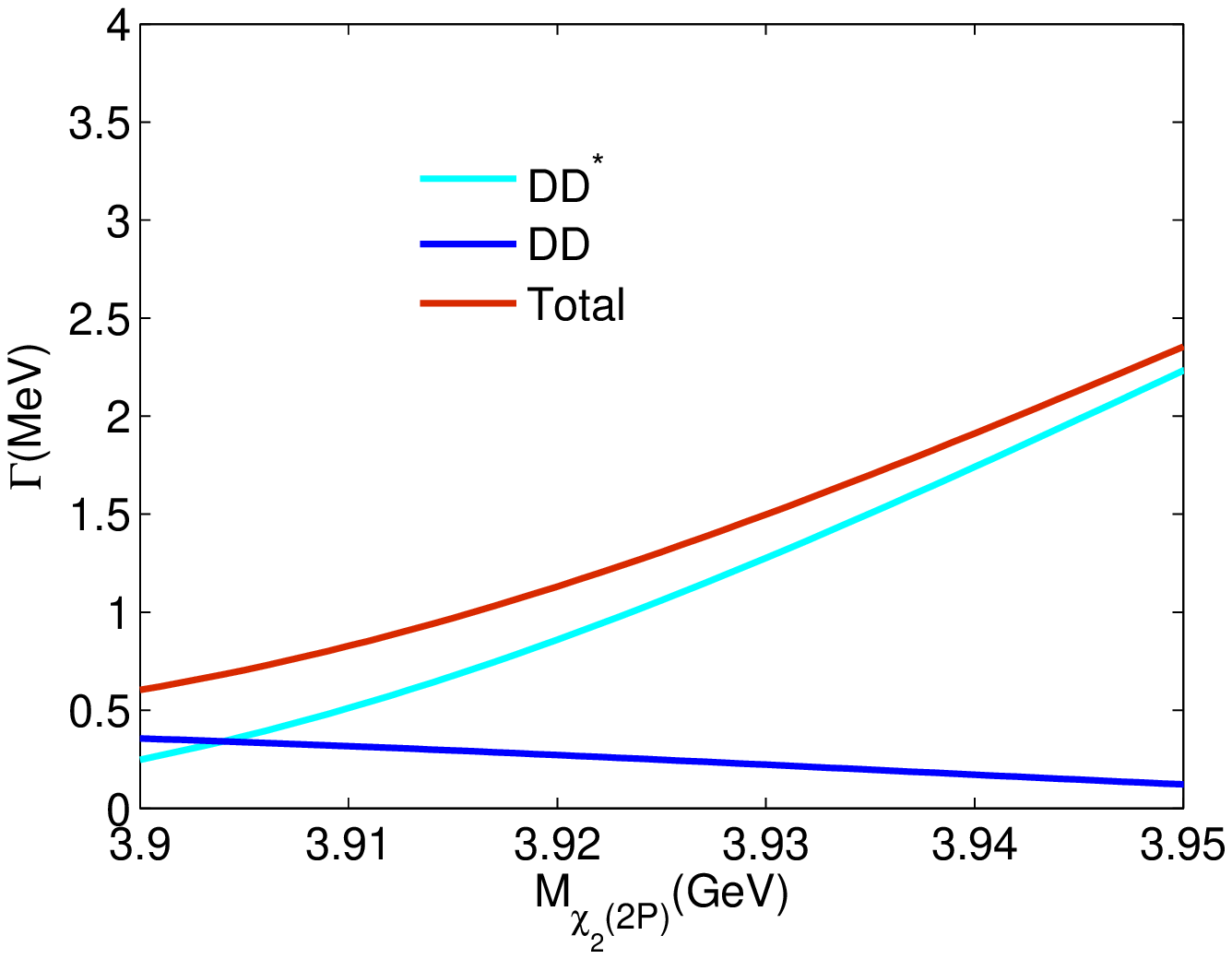}
\caption{The relation of the strong decay width with the mass of
$\chi_{c2}(2^3P_{2})$.\label{your label}}
\end{minipage}
\end{figure}

Whether we consider $X^{*}(3860)$ as $0^{++}$ or $2^{++}$ charmonium
state, there is only one strong decay mode, $X(3860)\rightarrow
D\overline{D}$, where $D$ refers to either $D^{0}$ or $D^{+}$. From
Figures 2 and 3, we can clearly see the deference between the total
decay widths of these two states. Taking $R=2.3\sim2.5$GeV$^{-1}$
discussed above, the total strong decay width of $0^{++}$ state
ranges from $110$ to $180$MeV, which is compatible with the
experimental data in Ref.~\cite{Bell1}. The total decay width of
$2^{++}$ state, which ranges from $0.4\sim1.9$MeV, is much smaller
than the experimental data. That means, if we assume $X^{*}(3860)$
as the $0^{++}$ charmonium state, its dominant decay mode and total
decay width is consistent well with the experimental data. Thus, our
present work support $X^{*}(3860)$ as the $0^{++}(\chi_{c0})$
charmonium state.

Considering $X(3915)$ as the $0^{++}$ and $2^{++}$ charmonium
respectively, we also observe different strong decay behaviors from
Figures 4 and 5. For the $0^{++}$ state, the total strong decay
width ranges from $159$ to $220$MeV, which dominantly decays into
$D\overline{D}$. Not only its total decay width but also the
dominant decay channel is inconsistent with the experimental data in
Ref.~\cite{Lees1}(See Table I). This means that the $X(3915)$ was
assumed to be the charmonium state $0^{++}$ is disfavored. If it is
treated as the $2^{++}$ charmonium, its decay behavior is very
similar with that of $X(3930)$(See Figures 5 and 6). They both decay
into $D\overline{D}$ and $D\overline{D}^{*}$ with the total decay
width ranging from $1.0$ to $3.0$MeV. In addition, these values of
the decay widths fall in the range of the experimental data. Thus,
it seems reasonable to assign both $X(3915)$ and $X(3930)$ to be the
the $2^{++}(\chi_{c2})$ charmonium state. If this conclusion is
true, the mass of the $\chi_{c2}$ charmonium state has erros. So, we
plot the relations of the strong decay widths on the masses of
$\chi_{c2}(2^{++})$ in Figure 7, which will be helpful to determine
its mass in the experimental and theoretical explorations in the
future.

Since the decay width of $\chi_{c2}\rightarrow D\overline{D}$ is
larger than $1$MeV, it should be observable in experiments for both
$X(3930)$ and $X(3915)$. However, it was reported by both Bell and
Babar Collaborions that the $X(3930)$ and $X(3915)$ were observed in
two different decay channels, $X(3930)\rightarrow D\overline{D}$ and
$X(3915)\rightarrow J/\psi\omega$. A reanalysis presented in
Ref.~\cite{ZhouZY} shows that if helicity-2 dominance assumption is
abandoned and a sizable helicity-0 component is allowed, the decay
process $X(3915)\rightarrow D\overline{D}$  may be reproduced in the
experimental data. But the large helicity-0 contribution means that
$X(3930)/X(3915)$ might not be a pure $c\overline{c}$ charmonium
state.

Since $X(3872)$ was observed, there have accumulated abundant
experimental information, which can be seen in Table I. Belle
experiment indicated $B(X(3872)\rightarrow
D^{0}\overline{D}^{0}\pi^{0}K^{+})=9.4^{+3.6}_{-4.3}B(X(3872)\rightarrow
J/\psi\pi^{+}\pi^{-}K^{+})$~\cite{Gokhroo}. Based on these
experimental data, we can draw a conclusion that
$\overline{D}^{*0}D^{0}$ is the dominant decay of $X(3872)$.
Although the underlying structure of this state is very
controversial, there is no doubt that its quantum number is
$1^{++}$. As a charmonium state $\chi_{c1}(1^{++})$, we show the
dependence of the strong decay width on the scale parameter $R$ in
Figure 8. Taking $R=2.3-2.5$GeV$^{-1}$, the decay width of the
inclusive decay channel $\overline{D}^{*0}D^{0}$ ranges from $0.2$
to $1.0$MeV, which falls in the range of the experimental data in
Table I and is also consistent with the conclusion of
$\overline{D}^{*0}D^{0}$ being the dominant decay mode. Thus, our
present work implies that $X(3872)$ is assigned to be the
$\chi_{c1}(1^{++})$ charmonium state is reasonable.

\begin{figure}[h]
\begin{minipage}[t]{0.45\linewidth}
\centering
\includegraphics[height=5cm,width=7cm]{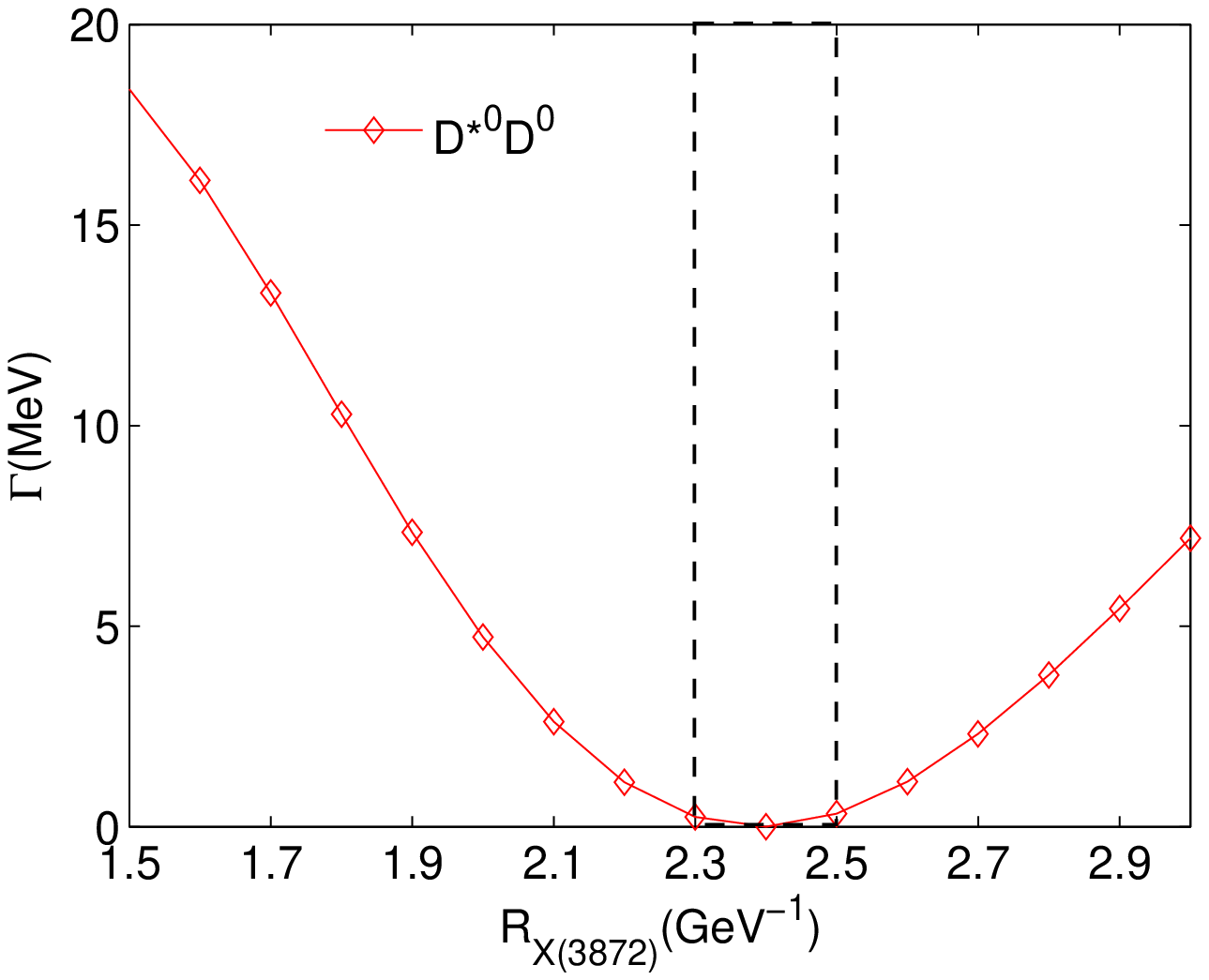}
\caption{The strong decay of $X(3872)$ as the $1^{++}(2^{3}P_{1})$
state on scale parameter $R_{X(3872)}$.\label{your label}}
\end{minipage}
\hfill
\begin{minipage}[t]{0.45\linewidth}
\centering
\includegraphics[height=5cm,width=7cm]{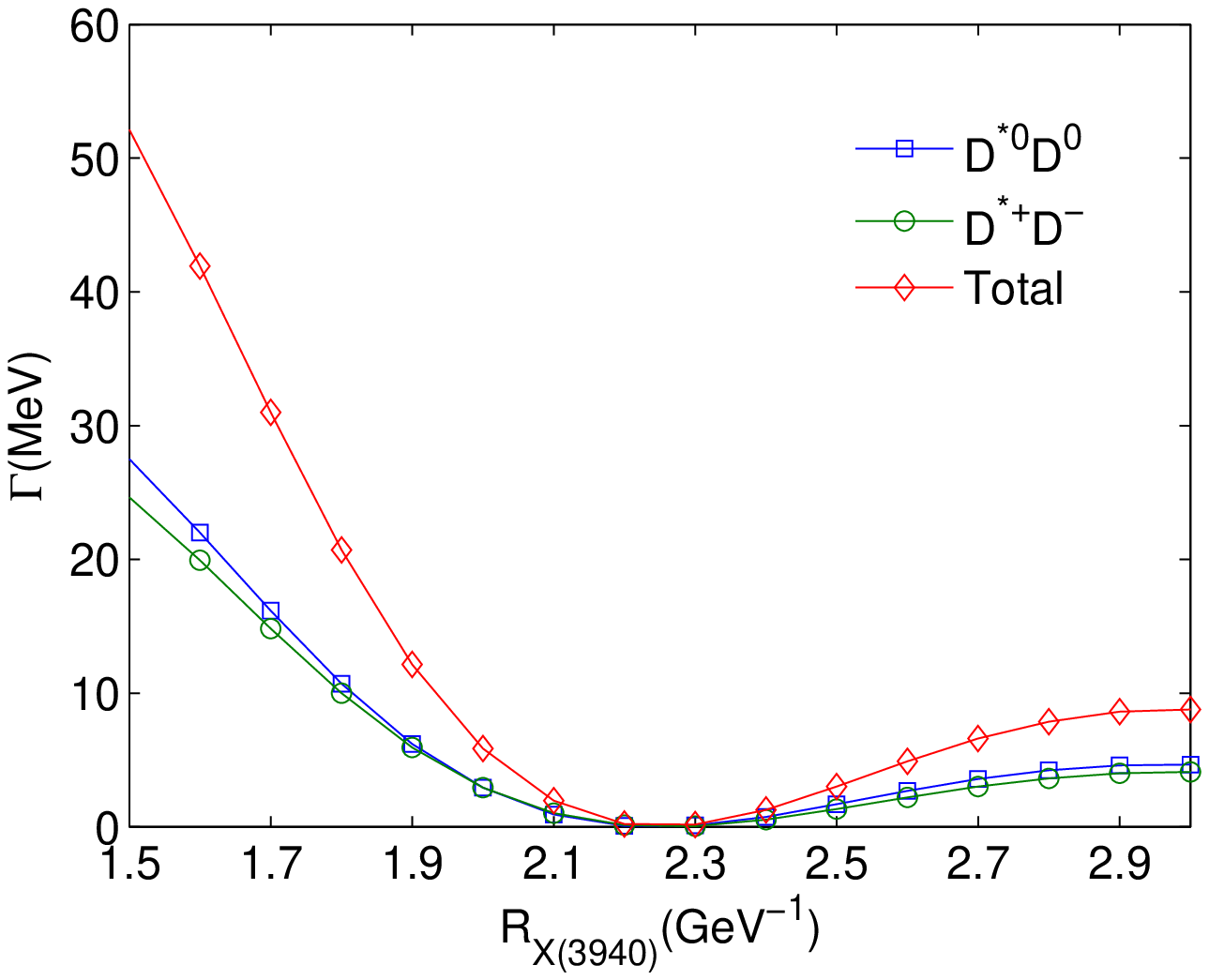}
\caption{The strong decay of $X(3940)$ as the $0^{-+}(3^{1}S_{0})$
state on scale parameter $R_{X(3940)}$.\label{your label}}
\end{minipage}
\end{figure}

Finally, we can see in Figure 9 that, as a $3^{1}S_{0}$ charmonium
state, $X(3940)$ can decay into $D^{*0}D^{0}$ and $D^{*+}D^{-}$
final states. This result is consistent with the experiments, where
$X(3940)$ was truly observed from the inclusive process
$e^{+}e^{-}\rightarrow J/\psi \overline{D}^{*}D$. However, it can
also be seen from Figure 9 that the maximum of the total decay width
can only reach up to $10$MeV if $R$ is changed from $2.0$ to
$3.0$GeV$^{-1}$. The predicted decay width in experiments is
$\Gamma=37^{+26}_{-15}\pm8$MeV which is much larger than this value.
This comparison indicates that $3^{1}S_{0}$ charmonium state might
not be a good candidate for the $X(3940)$.

\begin{large}
\textbf{4 Conclusion}
\end{large}

In summary, by considering both $X^{*}(3860)$ and $X(3915)$ as
$0^{++}$ and $2^{++}$ charmonium states, $X(3872)$, $X(3930)$,
$X(3940)$ as $1^{++}$, $2^{++}$ and $0^{-+}$ charmonium separately,
we study its two-body open charm strong decay behaviors by the
$^{3}P_{0}$ decay model. According to comparing our results with the
experimental data, we find that $X^{*}(3860)$ and $X(3872)$ can be
explained to be the $\chi_{c0}(2^{3}P_{0})$ and
$\chi_{c1}(2^{3}P_{1})$ charmonium state separately. The decay width
of $X(3940)$ is inconsistent with the experimental data if it is
supposed to be a $3^{1}S_{0}$ charmonium state. Thus, $3^{1}S_{0}$
charmonium state can be ruled out at prsent as a candidate for
$X(3940)$. Treated as a $0^{++}$ charmonium, the decay behavior of
$X(3915)$ is contradictory to experimental data. This indicates that
$X(3915)$ is unlikely to be a $0^{++}$ charmonium state. Supposed as
a $2^{++}$ charmonium, the decay behavior of $X(3915)$ is consistent
with not only the experimental data by also that of $X(3930)$. Thus,
we tentatively assign these two states as the same charmonium
$\chi_{c2}$. According to a reanalysis of the experimental data,
Zhou~\cite{ZhouZY} also suggested them to be the same state
$2^{++}$, but with a significant non-$c\overline{c}$ component. As a
result, the structure of $X(3915)/X(3930)$ needs to be further
studied according to more experimental and theoretical explorations.

\begin{large}
\textbf{Acknowledgment}
\end{large}

This work has been supported by the Fundamental Research Funds for
the Central Universities, Grant Number $2016MS133$.

\end{document}